\documentclass[fleqn,usenatbib]{mnras}

\usepackage{newtxtext,newtxmath}

\usepackage[T1]{fontenc}
\usepackage{ae,aecompl}

\usepackage{graphicx}	
\usepackage{amsmath}	
\usepackage{amssymb}	
\usepackage{comment}

\usepackage{soul}	
\usepackage{xcolor}

\title[Gamma--ray signal from SNe]{Time--dependent high-energy gamma--ray signal from accelerated particles in core-collapse supernovae: the case of SN~1993J}

\author[Cristofari et al.]{
P. Cristofari,$^{1,2}$\thanks{E-mail: pierre.cristofari@gssi.it}
M. Renaud,$^{3}$
A. Marcowith,$^{3}$ 
V.V. Dwarkadas,$^{4}$ and
V. Tatischeff~$^{5}$
\\
$^{1}$Gran Sasso Science Institute, via F. Crispi 7--67100, L'Aquila, Italy \\
$^{2}$INFN/Laboratori Nazionali del Gran Sasso, via G. Acitelli 22, Assergi (AQ), Italy \\
$^{3}$ Laboratoire Univers et Particules de Montpellier (LUPM), Universit\'e de Montpellier, CNRS/IN2P3, CC72, place Eug\`ene Bataillon, \\ F-34095 Montpellier Cedex 5, France \\
$^{4}$ Department of Astronomy and Astrophysics, University of Chicago, 5640 S Ellis Ave, Chicago, IL 60637, USA \\
$^{5}$ Universit\'e Paris-Saclay, CNRS/IN2P3, IJCLab, 91405 Orsay, France
}

\date{Accepted XXX. Received YYY; in original form ZZZ}

\pubyear{2015}

\begin{document}
\label{firstpage}
\pagerange{\pageref{firstpage}--\pageref{lastpage}}

\maketitle

\begin{abstract}
Some core--collapse supernovae are likely to be efficient cosmic--ray accelerators up to the PeV range, and therefore, to potentially play an important role in the overall Galactic cosmic--ray population. The TeV gamma--ray domain can be used to study particle acceleration in the multi--TeV and PeV range. This  motivates the study of the detectability of such supernovae  by current and future gamma--ray facilities.
The gamma--ray emission of core--collapse supernovae  strongly depends on the level of the two-photon annihilation process: high--energy gamma--ray photons emitted at the expanding shock wave following the supernova explosion can interact with soft photons from the supernova photosphere through the pair production channel, thereby strongly suppressing the flux of gamma rays leaving the system. In the case of SN~1993J, whose photospheric and shock--related parameters are well measured, we calculate the temporal evolution of the expected gamma--ray attenuation by accounting for the temporal and geometrical effects. We find the attenuation to be of about $10$ orders of magnitude in the first few days after the SN explosion. 
The probability of detection of a supernova similar to SN~1993J with the Cherenkov Telescope Array is highest if observations are performed either earlier than 1 day, or later than 10 days after the explosion, when the gamma--ray attenuation decreases to about $2$ orders of magnitude. 
\end{abstract}



\section{Introduction}
Gamma--ray observations of supernova remnants (SNRs)  have shown that these objects can efficiently accelerate particles up to the very--high--energy (TeV) domain. The detection and study of several SNRs~\citep{albert2007,acciari2009,HESSSNR} in the gamma--ray domain have helped us gain valuable understanding on the mechanisms involved at SNR shocks, and the role played by SNRs in the production of Galactic cosmic rays (CRs)~\citep[see e.g.][for reviews on the topic]{drury2012,blasi2013,amato2014}.

To date, two observational results are still fundamentally challenging the hypothesis that SNRs are the main sources of Galactic CRs. First, the fact that the spectra of all SNRs  detected in the gamma--ray domain are steeper than $E^{-2}$~\citep{acero2015}, which is the expected spectrum in the standard diffusive shock acceleration mechanism~\citep{axford1977,krymskii1977,bell1978,blandford1978}. This issue has been addressed by several theoretical works, investigating different hypotheses, as for instance  a steepening of the CR spectrum due to magnetic field amplification \citep{bell2019}, the geometrical effects of the expansion of a shock in a structured magnetic fields~\citep{malkov2019},  or the effects of particle escape  from the SNR~\citep{celli2019}. 

 The second issue with the SNR paradigm for the origin of Galactic CRs is the fact that all detected SNR shells to date have been shown to not be \textit{pevatrons}~\citep{gabici2019}, i.e. to be unable to accelerate protons up to PeV energies at their current evolutionary stage, which is required for sources in order to reach the \textit{knee} of the CR spectrum, located at $\sim 3$~PeV for protons. 
 
The possibility for core--collapse supernovae (CCSNe) to accelerate TeV-PeV particles has been addressed by several groups~\citep[see e.g.][]{tatischeff2009,bell2013,schure2014,marcowith2014,murase2014,cardillo2015,Giacinti15, zirakashvili2016,petropoulou17, bykov2918,marcowith2018, murase2019,fang2019} and reviews \citep{bykov18, tamborra18}.
 The fast shock resulting from the stellar explosion expanding in the dense wind of a red supergiant (RSG) progenitor was shown to be able to excite non--resonant streaming instabilities, thus allowing for efficient magnetic field amplification and particle acceleration into the PeV range in the early stages (typically in the first tens of days) after the SN explosion~\citep{tatischeff2009,marcowith2018}. This result supports the idea that CCSNe might be pevatrons, leading to the intriguing possibility that the resulting gamma--ray emission may be detectable by future ground--based gamma--ray facilities such as the Cherenkov Telescope Array (CTA)~\citep{CTAscience}. However, till now, no detection has yet been reported at GeV \citep{Fermi15} and TeV \citep{Hess19} gamma--ray energies.

Although simple energetic considerations indicate that the TeV gamma--ray luminosity from CCSNe might be sufficient for a detection \citep{kirk1995}, several effects could degrade the emitted gamma--ray signal. The most significant of these effects is the expected attenuation of the high--energy gamma--ray flux in the radiation field produced by the SN photosphere: very--high--energy gamma rays, typically in the TeV range, can interact with the dense photon field, typically in the eV range, through pair production:~$\gamma \gamma \rightarrow e^+ e^-$~\citep{gould1966,gould1967}. This effect has been studied and discussed in the context of other astrophysical objects, such as gamma--ray binaries~\citep{dubus2006}, and  CCSNe~\citep{marcowith2014} at a preliminary level. In most of the studies dealing with the detectability of CCSNe as cosmic-ray sources, the opacity due to this two-photon annihilation process has been treated as isotropic~\citep{tatischeff2009,murase2014,wang2019}. As we show below, however, the effect is distinctly non-isotropic, due to the dis-similar evolution of the radius of the SN photosphere and  that of the outer shock. In effect, at early timescales, the soft photons originating from the SN photosphere are produced at small distances from the gamma--ray source. 
Qualitatively, in the observer rest frame, their distribution would appear close to isotropic. But in time, the soft photons are produced increasingly further behind the shock, becoming  at some point hardly capable of catching the gamma--ray photons to interact with, thus producing a drop in the $\gamma \gamma$ absorption. 


To our knowledge, an extensive calculation of the $\gamma \gamma$ absorption, including both the geometrical effects and the temporal evolution of the expanding shock and of the SN photosphere, has not yet been carried out. This is the purpose of this paper, in which we present the calculations of the time--dependent $\gamma\gamma$ opacity (Sec.~\ref{sec:absorption}) and demonstrate its effect on the early gamma--ray emission from CCSNe. We illustrate our results using the well--known supernova SN~1993J, which encourages early targeted observations of CCSNe with CTA (Sec.~\ref{sec:gamma}). Our conclusions are presented in Sec.~\ref{sec:conclusions}.

\section{Time dependent gamma-gamma absorption}
\label{sec:absorption}

\begin{figure}
\includegraphics[width=.5\textwidth]{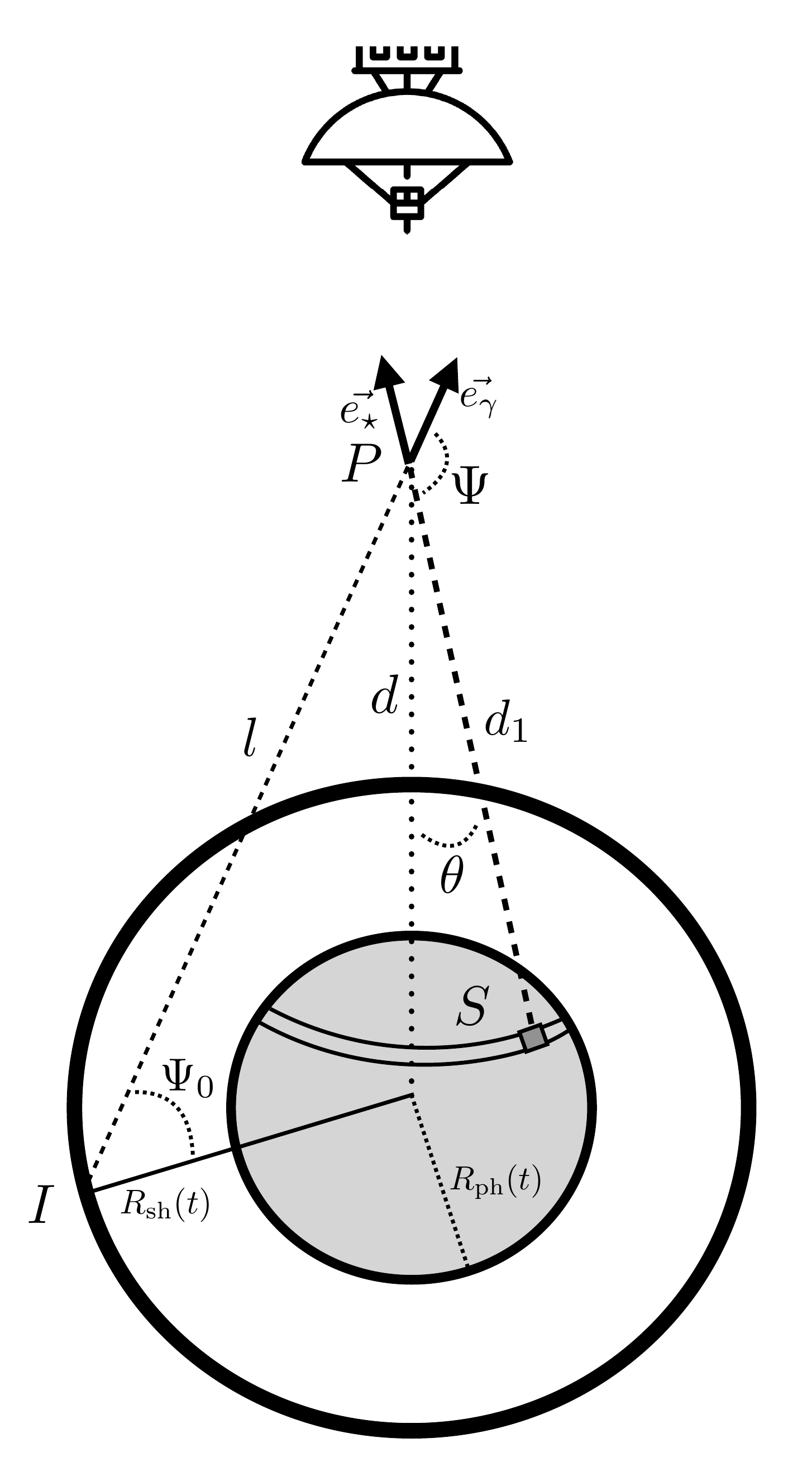}
\caption{Geometry considered in this work with a forward shock at $R_{\rm sh}(t)$ and a photosphere at $R_{\rm ph}(t)$. The interaction between a gamma--ray photon emitted at (I) and a soft photon emitted at (S) (surface of the photosphere) occurs at (P). The other parameters defining the system are described in the text.}
\label{fig:schema}
\end{figure}

The gamma--gamma opacity calculation is primarily a geometrical problem, where the gamma rays produced at the SN forward shock interact with lower energy photons from the SN photosphere. 
In this work, we follow the approach proposed in~\citet{dubus2006}~(D06), and adopt similar notations to carry out the opacity calculation. The main difference in the present work, compared to D06, lies in the inclusion of temporal effects: the position of the source of gamma rays (SN shock radius~$R_{\rm sh}$) and the position of the source of lower energy photons (SN photosphere~$R_{\rm ph}$) are both functions of time. In addition, the emission times of gamma--ray photons and of soft photons must be taken into account to determine which photons can interact at a given location denoted (P), after travelling from their emission locations denoted (I) and (S), as shown in Fig.~\ref{fig:schema}.

The calculation of the total opacity for a gamma--ray photon at a given energy $E$ requires a sextuple integration on: $\epsilon$, the soft photon energy; $\theta$ the polar angle at (P) between the center of the photosphere and the position (S) of the soft photon emitting region; $\phi$, the corresponding azimuthal angle; $l$, the distance between (I) the gamma--ray emitting region and (P) the interaction point ; $\psi_0$, the angle of the emitted gamma--ray photon at the interaction point; and $t$, the time after the SN explosion. 

\begin{equation}
\label{eq:tau_general}
\begin{aligned}
\tau_{\gamma \gamma}(E)=& \int_{0}^{t} \text{d}t' \int_{\psi_{0, \rm min}}^\pi \text{d}\psi_0 \int_{0}^{+\infty} \text{d}l  \\ 
& \int_{c_{\rm min}}^{1} \text{d} \cos \theta \int_{0}^{2 \pi}\text{d}\phi \int_{\epsilon_{\rm min}}^{+\infty} \text{d}\epsilon \frac{\text{d}\tau_{\gamma \gamma}}{\text{d}\epsilon \text{d}\Omega \text{d}l} \ ,
\end{aligned}
\end{equation}
where the differential absorption opacity reads~\citep{gould1966,gould1967}: 
\begin{equation}
\text{d}\tau_{\gamma\gamma}= \left(1- {\bf e_{\gamma} e_{\star}} \right)n_{\epsilon} \sigma_{\gamma \gamma} \text{d}\epsilon \text{d}\Omega \text{d}l~, 
\end{equation}
with ${\bf e_{\gamma}}$ and ${\bf e_{\star}}$ denoting the direction of the  interacting gamma--ray photon and soft photon respectively. The cross section $\sigma_{\gamma \gamma}$ for the pair production process $\gamma + \gamma \rightarrow e^+ + e^-$ is derived in~\citet{gould1967}, $\text{d}\Omega= \sin \theta \text{d}\phi \text{d}\theta$ is the solid angle of the surface emitting the photons of energy $\epsilon$ and $n_{\epsilon}$ is the radiation density.  Eq.~(\ref{eq:tau_general}) is a generalization of Eq.~(A.8) of D06 to take into account the temporal effects, which requires the calculation of two more integrals on the time $t$ and emission angle $\psi_0$.

The photospheric photon density is assumed to follow a blackbody distribution with a time-dependent temperature $T_{\rm ph}(t)$: 
\begin{equation}
    n (\epsilon, t)= \frac{2 \epsilon^2}{h^3 c^3} \frac{1}{\exp (\epsilon/kT_{\rm ph}(t))-1} ~~\text{cm}^{-3} ~\text{erg}^{-1} ~\text{sr}^{-1}.
\end{equation}

Accounting for temporal effects requires the calculation of the emission time of a gamma--ray photon and of a soft photon which interact at (P). Let us define the interaction time of a gamma--ray photon:
\begin{equation}
t_{I}= t+ l/c~.
\end{equation}
This time is a function of $l$, and is redefined at each $\text{d}l$. Integrating over $\cos\theta$, we define the time at which the soft photon interacting with the gamma--ray photon has been emitted:
\begin{equation}
\label{eq:t_s}
t_{S}=t_I - d_1/c~,
\end{equation}
where $d_1$, the distance travelled by the soft photon, depends on $R_{\rm sh}(t)$, $d$, and $\cos\theta$.



The lower limit of the integral on $\cos \theta$, $c_{\rm min}(t,\psi_0)$, is a crucial quantity of the problem and requires a careful derivation. It is given by the maximum possible value of $\theta$, which depends on the time-dependent photospheric radius and the soft photon emission time. It is related to the distance $d_1$ defined in Eq.~\eqref{eq:t_s}, which satisfies the relation:

\begin{equation}
d_1^2-2 d_1 d \cos \theta +d^2-R^2_{\rm ph}(t=d_1/c)=0~.
\end{equation}

The corresponding photosheric radius $R_{\star}$ and emission time $T_{\rm ph}$ needed to calculate $\theta_{\rm max}$ are therefore: 
\begin{equation}
\begin{aligned}
T_{\rm ph} &= t_I-d_1/c~, \\
R_{\star} &= d_1 \sin(\theta_{\rm max})~,\\
\end{aligned}
\end{equation}
and
\begin{equation}
c_{\rm min}(t,\psi_0)= \sqrt{1-\left(\frac{R_{\star}}{d_1}\right)^2}~.
\end{equation}

 Another crucial quantity of the problem is the minimum emission angle $\psi_{0,\rm min}$, which corresponds to the limit below which gamma rays are absorbed because they have to cross the photosphere.  It is obtained at a given time $t$ from the condition $R_{\rm sh}(t) \sin(\psi_{0, \rm min}) = R_{\rm ph}(t_l)$, where we search for all times $t_l > t$:
\begin{equation}
\label{eq:psi0min}
\psi_{0,\rm min} (t)= \arcsin \left(\frac{\text{Max}(R_{\rm ph}(t_l))_{t_l>t}}{R_{\rm sh}(t)} \right) ~.
\end{equation}
An alternative method to derive $\psi_{0, \rm min}$ is proposed in Appendix~\ref{app:A}. \\

Another time effect is due to the Doppler frequency shift of the photons between their emission site and the interaction point. The relative frequency shift relative to the emission frequency $\nu_{\rm em}$ is $\Delta \nu/\nu_{\rm em} \simeq \Delta V/c$, where $\Delta V$ is the forward shock speed $V_{\rm sh}$ for the gamma--ray photons (or the photospheric expansion speed $V_{\rm ph}$ for the soft photons) . In both cases the ratio of these speeds to the speed of light is $\ll 1$ except at the earliest times where $V_{\rm sh}$/c $\sim$ 0.1. Hence the Doppler shift is a small effect. We nevertheless account for it in the gamma-gamma opacity calculation. 

In practice, the calculation of $\tau_{\gamma \gamma}$ defined in Eq.~\eqref{eq:tau_general} is carried out defining grids for the several variables of the problem: $E, t, \psi_0, l, \theta, \phi$ and $\epsilon$, and evaluating the successive integrals using simple trapezoidal integration in logarithmic space, and linear space for $\cos \theta $, $\psi_0$ and $\phi$.
The calculation can be parallelized, for example by evaluating simultaneously the integrals on $\psi_0, l, \theta, \phi$ and $\epsilon$ for each points of the grids in $t$ and $E$. 

\section{Time--dependent gamma--ray signal}
\label{sec:gamma}
The time dependent gamma--gamma absorption effects described in Sec.~\ref{sec:absorption} are calculated for the case of the well-known supernova SN~1993J. SN~1993J, discovered in the Galaxy M81~(NGC~3031)~\citep{ripero1993}, is one of the brightest optically detected SNe in the northern hemisphere. It resulted from the collapse of a binary system with a progenitor mass in the range of 13--20~M$_{\odot}$~\citep{maund2004}. 
Early studies of SN~1993J estimated that the fast shock produced by the SN explosion, if evolving in a density profile $\propto r^{-3/2}$, at a distance of 3.63~Mpc, would result in a gamma--ray flux F(>1~TeV)$\approx  10^{-12}$~cm$^{-2}$s$^{-1}$~\citep{kirk1995}. These estimates were later shown to be an overestimation of the expected gamma--ray signal \citep{tatischeff2009} (T09), since the absorption of gamma rays by soft photons from the SN photosphere was not addressed by \citet{kirk1995}, and the density profile was shown to fall off faster than the former assumption, $\propto r^{-2}$ instead of  $\propto r^{-3/2}$. 

In this study, we assume that the unabsorbed gamma--ray flux from SN~1993J follows~(T09):
\begin{equation}
\begin{split}
\frac{\text{d} N}{\text{d} E}= 2 \times 10^{-12} \; \left(\frac{D}{3.63~\text{Mpc}} \right)^{-2} \left(\frac{\dot{M}_{\rm RSG}}{3.8 \times 10^{-5} M_{\odot}/\text{yr}} \right)^2 \\
\times \left(\frac{t}{\text{days}} \right)^{-1} \left( \frac{u_{\rm w}}{10 \;  \text{km/s}}\right)^{-2} \left( \frac{E}{1 \; \text{TeV}}\right)^{-2} \text{TeV}^{-1}~ \text{cm}^{-2}~ \text{s}^{-1} ~.
\label{eq:t09}
\end{split}
\end{equation}
under the assumption that the cosmic ray efficiency at the shock scales with time as $\xi_{\rm CR}=0.04 (t/1\text{day})^{0.17}$. Here,
$\dot{M}_{\rm RSG}$ is the mass-loss rate of the red supergiant progenitor, $u_{\rm w}$ the wind terminal velocity, and $D$ the distance to the object. Considering the low efficiency assumed here, it is reasonable to neglect the back reaction of accelerated particles on the shock structure which leads to a differential spectrum $\propto E^{-2}$ at relativistic energies. Although they are not a priori expected, deviations from an $\propto E^{-2}$ spectrum of accelerated particles are possible. Steeper spectra  would degrade the TeV gamma--ray signal produced at the shock~\citep{drury1994}. For instance, a spectrum scaling as $E^{-2.4}$ would reduce the unabsorbed integrated flux above 1 TeV by about 40\%. The impact of the wind parameters on the gamma-ray flux is apparent from Eq.~\eqref{eq:t09}.  In particular a variation of a factor 3 in the ratio $\dot{M_{\rm RSG}}/u_{\rm w}$ leads to a variation of about one order of magnitude of the unabsorbed spectrum. 

This work focuses on the test case of SN~1993J and we assume typical values for the input parameters to illustrate the effects of the $\gamma\gamma$ absorption on the detectability of this object. We consider that particles up to the $\sim$ PeV range  are accelerated from day one~\citep{marcowith2018}, and that these particles produce gamma rays in the energy range of interest up to $\sim 100$~TeV.

\begin{figure}
\includegraphics[width=.5\textwidth]{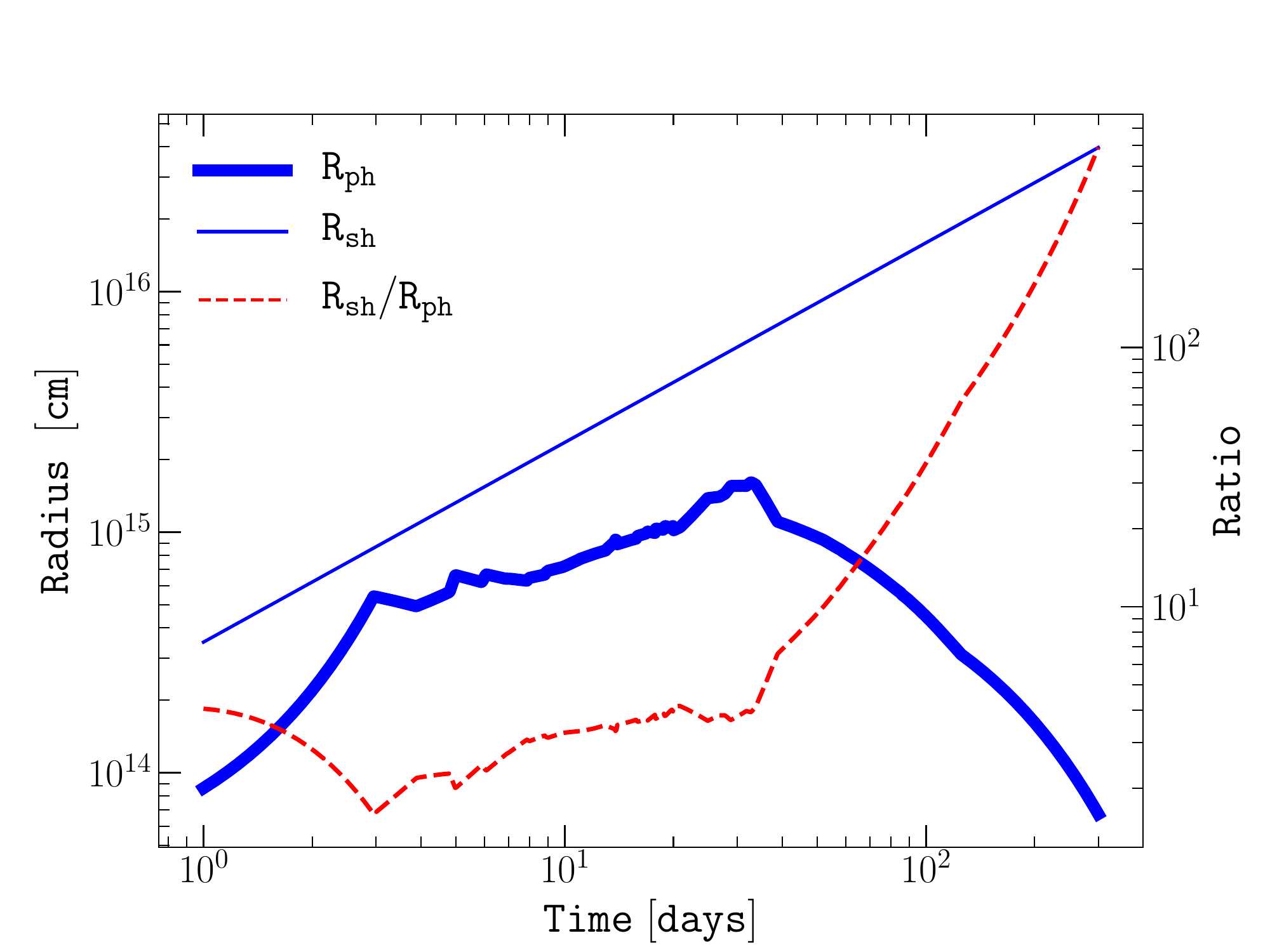}
\caption{Time evolution of the shock radius $R_{\rm sh}$ (blue solid thin line), the photospheric radius $R_{\rm ph}$ (blue solid thick line), and the ratio $R_{\rm sh}$/$R_{\rm ph}$ (red dashed line, right axis) in the case of SN~1993J.}
\label{fig:radii}
\end{figure}

\begin{figure}
\includegraphics[width=.5\textwidth]{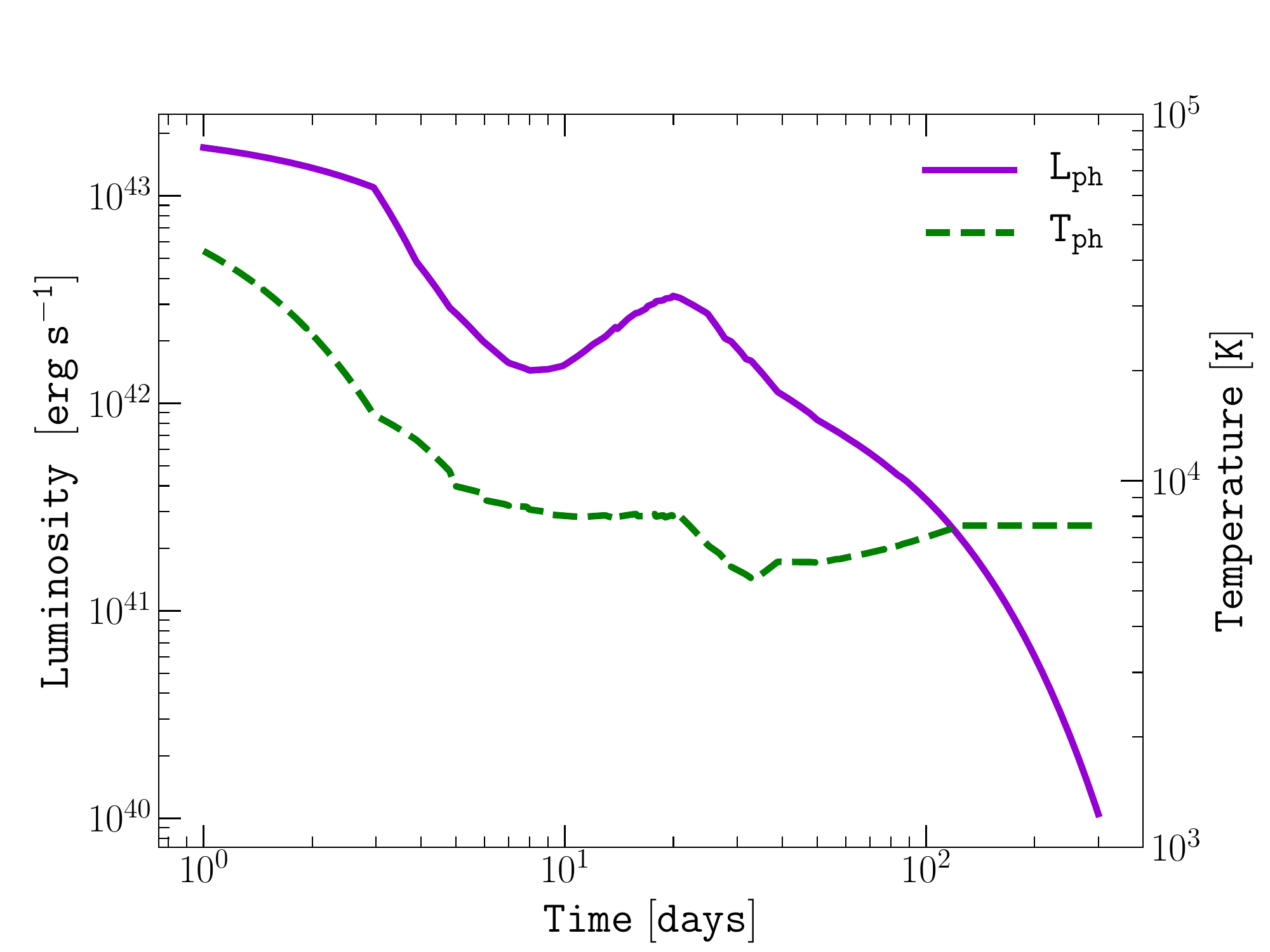}
\caption{Time evolution of the photosphere luminosity $L_{\rm ph}$ (violet solid line, left axis) and  the photosphere temperature $T_{\rm ph}$ (green dashed line, right axis)  in SN~1993J~\citep{lewis1994}. }
\label{fig:temperature}
\end{figure}

The expansion of the strong shock resulting from the SN explosion, in the dense RSG wind, can be described using self--similar solutions~\citep{chevalier1982,chevalier1994}. Following T09, we assume that the shock radius follows: 
\begin{equation}
R_{\rm sh}(t)= R_0 \times \left( \frac{t}{t_0} \right)^{m}
\end{equation}
with $R_0\approx 3.43 \times 10^{14}$~cm, $m=0.83$ and $t_0=1$~day. The evolution of the photospheric radius $R_{\rm ph}$ is obtained from optical observations~\citep{lewis1994}. The evolution of $R_{\rm sh}$ and $R_{\rm ph}$ is shown in Fig.~\ref{fig:radii}, while the evolution of the photospheric luminosity $L_{\rm ph}$ and temperature $T_{\rm ph}$ is displayed in Fig.~\ref{fig:temperature}.

\begin{figure}
\includegraphics[width=.5\textwidth]{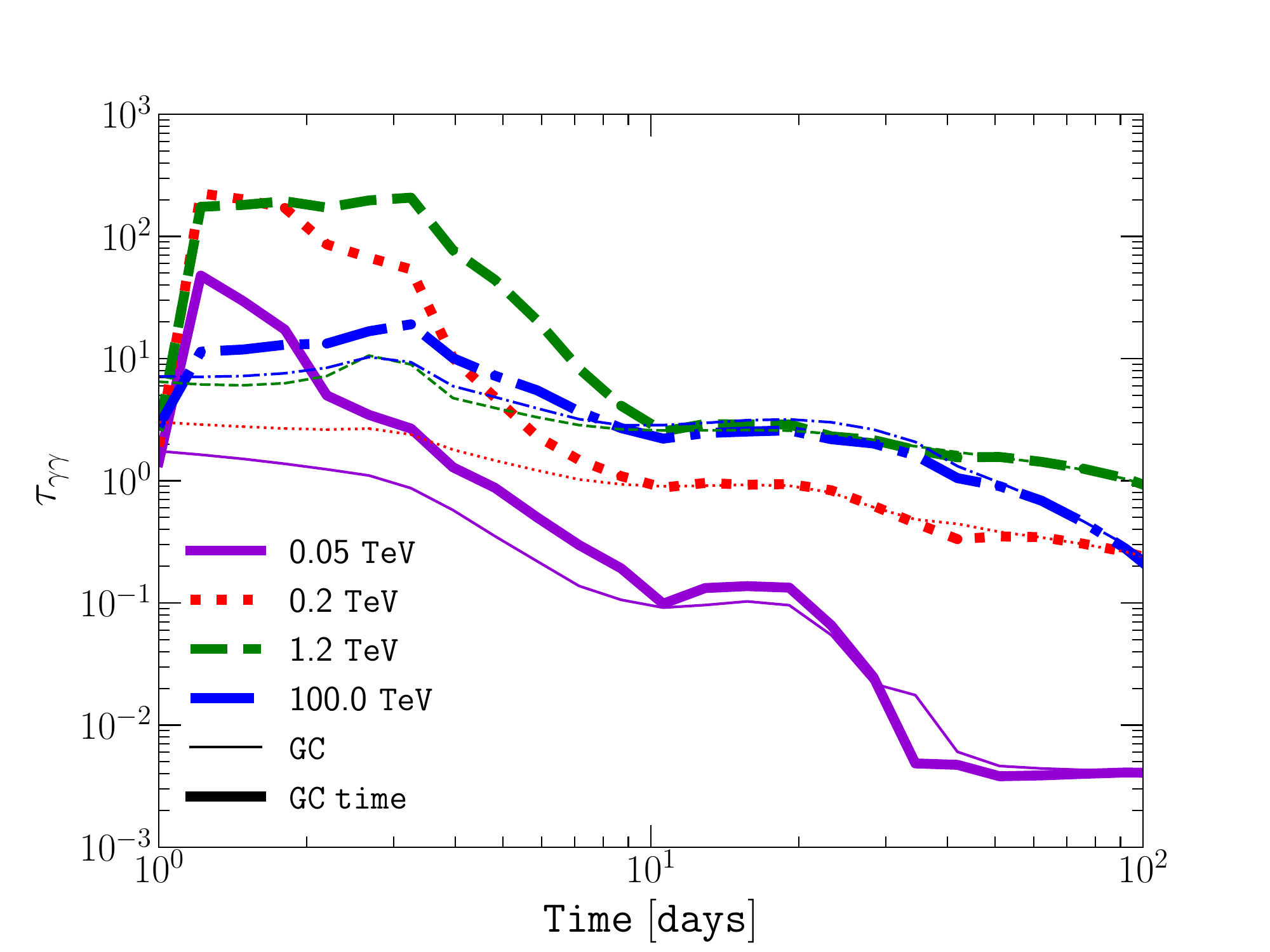}
\caption{Time evolution of the opacity $\tau_{\gamma \gamma}$ for gamma--rays of different energies. The general case with temporal effects (GC time, thick lines) and without temporal effects (GC, thin lines) are shown.}
\label{fig:opacity}
\end{figure}

We calculate the gamma--gamma absorption for four different scenarios: (1) Point source photosphere without temporal effects  (PS); (2) Point source photosphere including temporal effects (PS time); (3) General case without temporal effects  (GC); (4) General case including temporal effects (GC time). 
 The second case has already been treated in previous works~\citep{marcowith2014}. The PS cases (1) and (2) are simplified versions of the calculation of the GC cases (3) and (4). Indeed, for a given (t,$\psi_0$) the lower bound integration over $\cos \theta$ in Eq.~\eqref{eq:tau_general} is $c_{\rm min}(t,\psi_0)= (1- R^2_{\star}/d_1^2)^{1/2}$. For a point source, $R_{\star}\ll d_1$ and  $c_{\rm min}(t,\psi_0) \rightarrow 1$, thereby simplifying Eq.~\eqref{eq:tau_general}. 

The temporal effects correspond to the delay due to the different emission times of the gamma--ray and soft photons, and the duration of the travel from the emission points (I and S respectively) to the interaction point (P). Fig.~\ref{fig:opacity} illustrates the temporal evolution of the opacity $\tau_{\gamma \gamma}$ for different energies of the incident gamma--ray photons in the GC, with and without the temporal effects. The difference between the two cases is substantial up to $\approx 10$~days, and becomes negligible after $\approx 40$~days when $R_{\rm sh}/R_{\rm ph} \gtrsim 4$. 

\begin{figure}
\includegraphics[width=.5\textwidth]{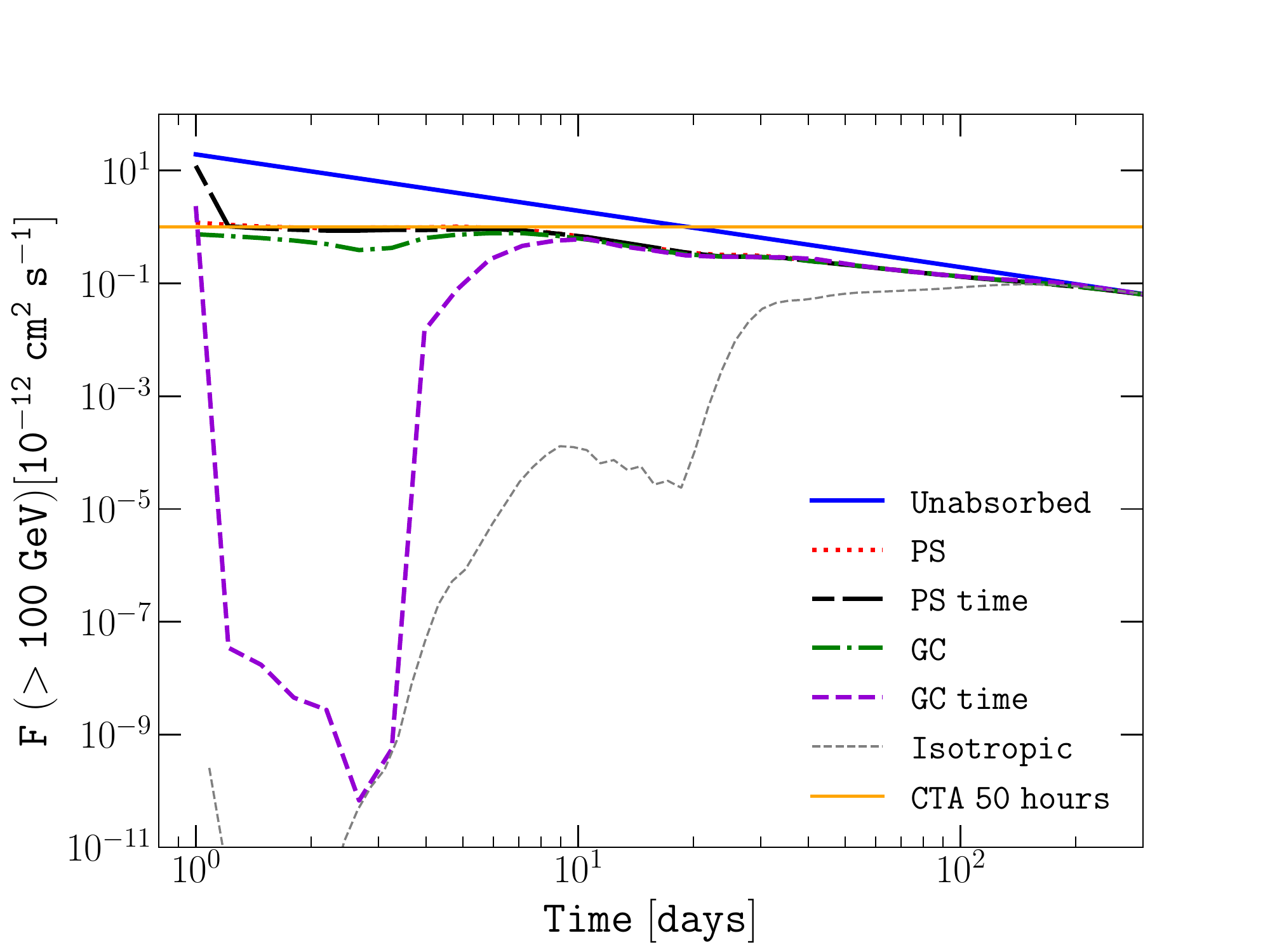}
\caption{Time evolution of the integrated flux above 100~GeV from SN~1993J. Six cases are shown: unabsorbed (blue solid, see Eq.~\eqref{eq:t09}), point source (PS, red dotted), point source with time effects (PS time, black dot--long dashed), general case (GC, green dot--dashed),  general case with time effects (GC time, purple dashed), and the isotropic calculation (grey thin dotted)~\citep{aharonian2008}. The typical sensitivity of CTA in 50 hours at energies above 100 GeV~\citep{fioretti2016} is shown as the horizontal orange line.}
\label{fig:100GeV}
\end{figure}

\begin{figure}
\includegraphics[width=.5\textwidth]{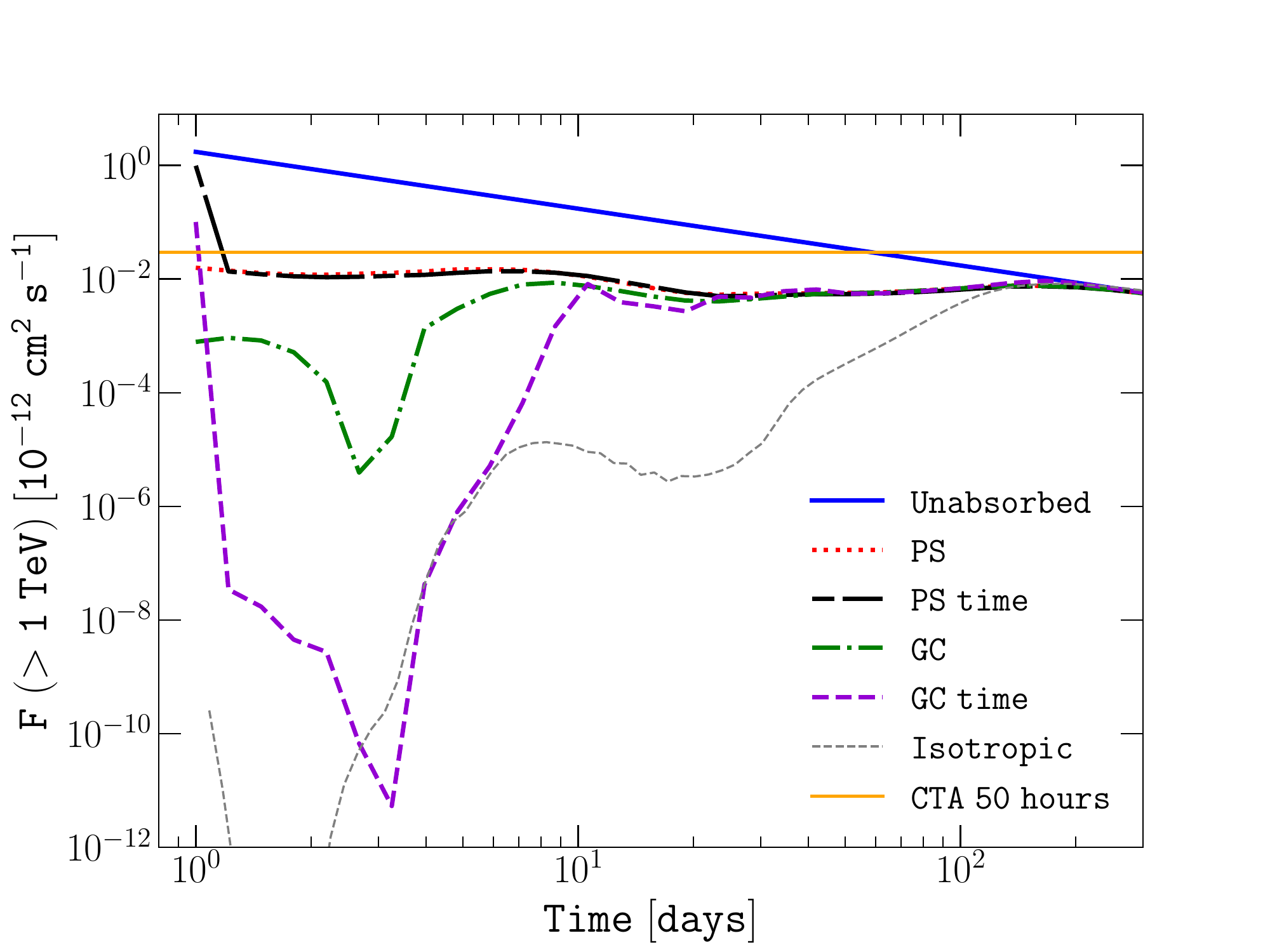}
\caption{Time evolution of the integrated flux above 1~TeV from SN~1993J. Legend is similar to Fig.~\ref{fig:100GeV}.}
\label{fig:1TeV}
\end{figure}

 We illustrate these effects for the four cases described above, by computing the time evolution of the integrated flux above 100~GeV and above 1~TeV as shown in Fig.~\ref{fig:100GeV} and Fig.~\ref{fig:1TeV}, respectively. We see in Fig.~\ref{fig:1TeV} that in the general case, the attenuation increases by up to $\approx 9$~orders of magnitude (and $\approx 5$~orders of magnitude without time effects) compared to the point source case. The attenuation is maximal at $\approx 3$~days where the difference between $R_{\rm sh}$ and $R_{\rm ph}$ is minimal. After $\gtrsim 10$~days, the calculation for the general cases coincides with the results obtained in the point source cases. Indeed, as mentioned in D06, the point source approximation accurately describes the system when $R_{\rm sh} \gtrsim 4 R_{\rm ph}$.

We additionally show the result of the attenuation calculation in the isotropic case, as in~\citet{aharonian2008} and T09, in order to illustrate that our results lead to more optimistic gamma--ray flux than these previous estimates.

\begin{figure}
\includegraphics[width=.5\textwidth]{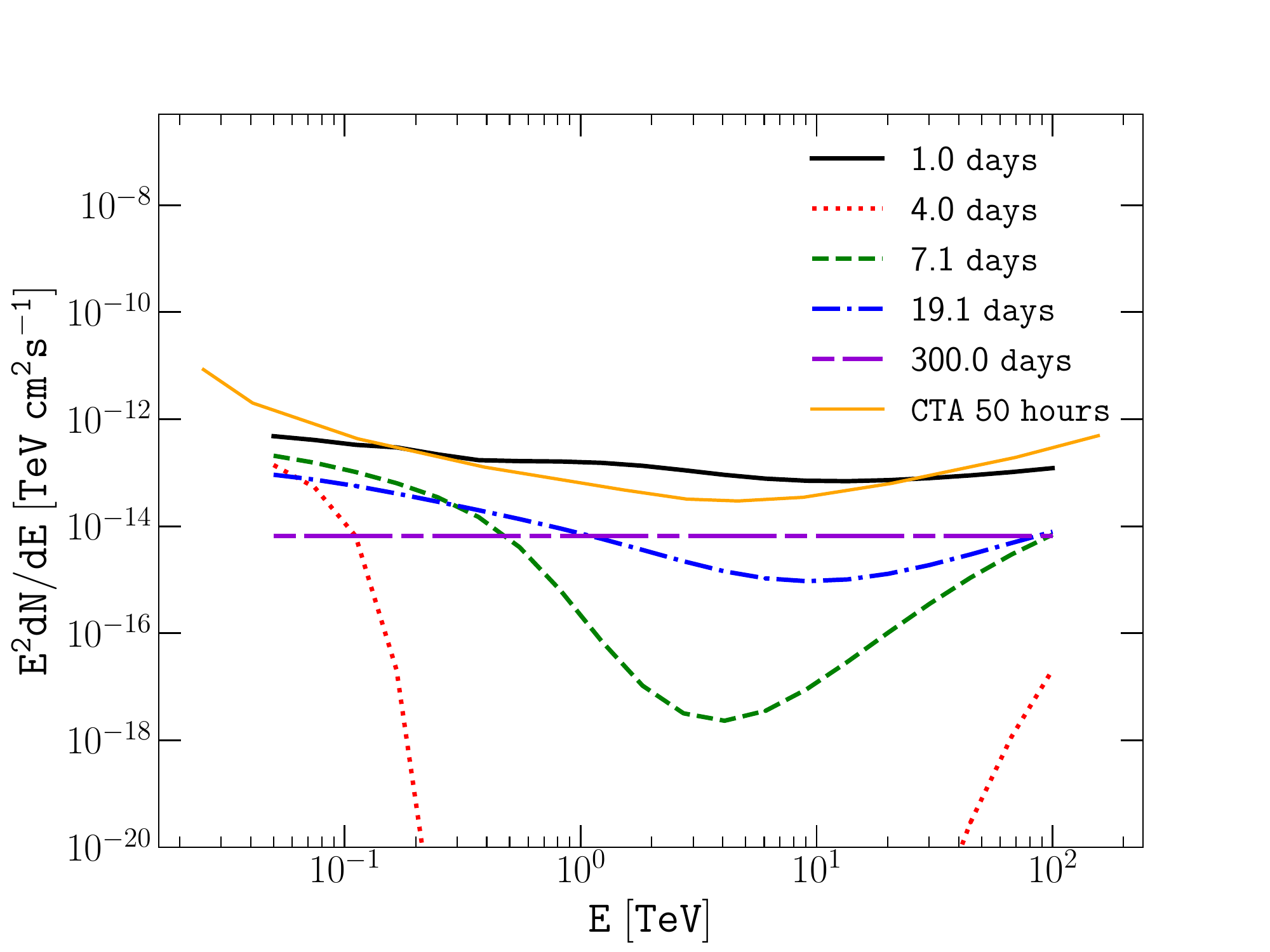}
\caption{Differential gamma--ray spectrum of SN~1993J at 1~day (solid black), 4~days (red dotted), 7.1~days (green dashed), 19.1~days (blue dot-dashed) and 300~days (purple dot-long dashed) after the SN explosion, in the general case taking into account the geometrical and temporal effects. The CTA sensitivity in 50 hours is also shown \citep{CTAscience}.}
\label{fig:diff}
\end{figure}

In the general case, taking into account the geometrical and temporal effects, we compute the differential spectrum at different times. The attenuation is maximal in the TeV range, as shown in Fig.~\ref{fig:diff}. After $\approx 10$ days, the flux around $\sim 1$~TeV and $\sim 100$~TeV becomes comparable to the unabsorbed one after 300 days, illustrating that the optimal windows for the detection of a gamma--ray signal in this case are either in the first day after the SN explosion, or after $\gtrsim 10$~days. As a guide for the reader,  we also show the typical integrated sensitivity of CTA for a point source observed for 50 hours~\citep{fioretti2016,CTAscience}. Although it appears that the obtained integrated fluxes above 100 GeV and above 1 TeV from SN~1993J are about one order of magnitude below the typical 50 hour sensitivity of CTA after $\gtrsim$ 10 days, a more careful analysis, taking into account the performance of the instrument, is needed. This would allow one to accurately quantify the detectability of such a SN~1993J-like SN event with CTA. 

 Pair production can also occur through the Bethe-Heitler process which involves a direct interaction of photons with a nuclei \citep{Blumenthal70}. The threshold for pair production $E_{\rm th}$ can be estimated equating $m_{\rm e} c^2 \sim k_{\rm B} T_{\rm ph}$. 
 The temporal evolution of $T_{\rm ph}$ in shown in Fig.~\ref{fig:temperature}. In the case of SN 1993J the photosphere temperature does not exceed $10^4$ K at early times, hence $E_{\rm th} \ge 10^6-10^7$ GeV.  At these energies, the secondary gamma-ray photons produced by neutral pion decay are only marginally compatible with the domain covered by CTA. Notice that the previous estimation is obtained assuming an isotropic soft photon distribution. Hotter photospheres produce lower thresholds and then may be relevant to explore for multi--hundred TeV sensitive telescopes like LHAASO~\citep{LHAASO}. A refined calculation including time-dependent and anisotropic effects will be proposed in a forthcoming work. A second Bethe-Heitler process is the direct interaction of gamma-ray photons with bounded electrons in atoms \citep{murase2014}. We have checked that the opacity for this process is never larger than one in the conditions that prevail for SN 1993J.
 
 Finally, let us consider other events besides SN~1993J. The scaling of the gamma-ray flux on the progenitor wind properties was shown above in Eq.~\eqref{eq:t09}. Further discussion of radio SNe expanding within their environments, the properties of the environment, and the maximum energies to which particles are accelerated, was done in \citet{marcowith2018}. A shock expanding at a larger velocity than that measured in SN~1993J (such as in Ib/c SNe \citep{smartt2009}), or in a denser circumstellar wind (such as in Type IIn SNe \citep{moriya2014}) would lead to a larger number of accelerated particles since the pressure in accelerated CRs scales as $P_{\rm CR} \approx \xi \rho_{\rm wind} V_{\rm sh}^2$, with $\xi$ the CR efficiency, $\rho_{\rm wind}$ the wind density and $V_{\rm sh}$ the shock velocity.   The denser wind would provide a larger amount of target material, and therefore an enhanced production of gamma--rays would be expected. However, without knowing the characteristics of the photosphere, it is a priori impossible to infer the level of attenuation due to gamma--gamma absorption, and therefore very hard to properly estimate the detectability of such object. The level of gamma--gamma absorption for different types of CCSNe will be studied in a future work.

\section{Conclusions}
\label{sec:conclusions}

We have calculated the time evolution of the gamma--gamma absorption at a CCSNe, and illustrated the level of attenuation of the gamma--ray flux in the case of SN~1993J. Our calculation takes into account time dependent effects, i.e. the evolution in time of the shock radius and the photospheric radius, and the emission times of the interacting gamma--ray and soft photons. The temporal evolution of the geometry makes this calculation difficult to handle. Our results indicate that it is crucial to take into account these temporal and geometrical effects to properly estimate the impact of the gamma--gamma absorption during the first weeks after the SN explosion. 

The calculation performed in the case of SN~1993J exemplifies the importance of the temporal evolution of both the shock and the photosphere. We obtain typically 8--10 orders of magnitude of difference in the gamma--ray flux, between a point source model and the general case in which temporal and geometrical effects are included. This attests to the importance of performing the calculation in the general case when the radii of the shock and of the photosphere are comparable. 




 Without a careful inclusion of the actual performance of an instrument optimized in the TeV range, such as CTA, it is difficult to firmly assert the detectability of SN~1993J--like event. However, our results suggests that a detection might be conceivable within 50~hours with CTA, either during the first day after the SN explosion, or after $\sim $10 days.

As illustrated on Fig.~\ref{fig:100GeV} and Fig.~\ref{fig:1TeV}, previous isotropic calculations~\citep{aharonian2008,tatischeff2009} were more pessimistic and suggested that in the first 300 days, the gamma--ray signal remains several orders of magnitude below the typical sensitivity of Imaging Atmospheric Cherenkov telescopes. Our results show that correctly taking into account the different emission times of the gamma-ray and soft photons in the opacity calculation reduces the likelihood of detection. However, the time period over which the absorption is significant is considerably shortened, to about 10 days from the previous 300 days. In this sense, our results, are more encouraging than these previous estimates.
A deeper study, based on the instrument response functions of the current and future IACTs would be very valuable.  

Moreover, other types of CCSNe exhibiting different shock-related and photospheric properties will be considered in a forthcoming work. In particular our work suggests that those SNe for which the shock radius would rapidly becomes larger than the photospheric radius might be promising targets for gamma--ray instruments optimized in the TeV range.

\section*{Acknowledgements}
The authors thank G. Dubus and A. Dundovic for helpful discussions and comments on this work. VVD's work is supported by NSF grant 1911061 awarded to the University of Chicago (PI V.~Dwarkadas). 

\bibliographystyle{mnras}
\bibliography{CCSNe} 

\begin{thebibliography}{}
\makeatletter
\relax
\def\mn@urlcharsother{\let\do\@makeother \do\$\do\&\do\#\do\^\do\_\do\%\do\~}
\def\mn@doi{\begingroup\mn@urlcharsother \@ifnextchar [ {\mn@doi@}
  {\mn@doi@[]}}
\def\mn@doi@[#1]#2{\def\@tempa{#1}\ifx\@tempa\@empty \href
  {http://dx.doi.org/#2} {doi:#2}\else \href {http://dx.doi.org/#2} {#1}\fi
  \endgroup}
\def\mn@eprint#1#2{\mn@eprint@#1:#2::\@nil}
\def\mn@eprint@arXiv#1{\href {http://arxiv.org/abs/#1} {{\tt arXiv:#1}}}
\def\mn@eprint@dblp#1{\href {http://dblp.uni-trier.de/rec/bibtex/#1.xml}
  {dblp:#1}}
\def\mn@eprint@#1:#2:#3:#4\@nil{\def\@tempa {#1}\def\@tempb {#2}\def\@tempc
  {#3}\ifx \@tempc \@empty \let \@tempc \@tempb \let \@tempb \@tempa \fi \ifx
  \@tempb \@empty \def\@tempb {arXiv}\fi \@ifundefined
  {mn@eprint@\@tempb}{\@tempb:\@tempc}{\expandafter \expandafter \csname
  mn@eprint@\@tempb\endcsname \expandafter{\@tempc}}}

\bibitem[\protect\citeauthoryear{{Acciari} et~al.,}{{Acciari}
  et~al.}{2009}]{acciari2009}
{Acciari} V.~A.,  et~al., 2009, \mn@doi [\apjl] {10.1088/0004-637X/698/2/L133},
  \href {https://ui.adsabs.harvard.edu/abs/2009ApJ...698L.133A} {698, L133}

\bibitem[\protect\citeauthoryear{{Acero}, {Lemoine-Goumard}, {Renaud},
  {Ballet}, {Hewitt}, {Rousseau}  \& {Tanaka}}{{Acero}
  et~al.}{2015}]{acero2015}
{Acero} F.,  {Lemoine-Goumard} M.,  {Renaud} M.,  {Ballet} J.,  {Hewitt} J.~W.,
   {Rousseau} R.,   {Tanaka} T.,  2015, \mn@doi [\aap]
  {10.1051/0004-6361/201525932}, \href
  {https://ui.adsabs.harvard.edu/abs/2015A&A...580A..74A} {580, A74}

\bibitem[\protect\citeauthoryear{{Ackermann} et~al.,}{{Ackermann}
  et~al.}{2015}]{Fermi15}
{Ackermann} M.,  et~al., 2015, \mn@doi [\apj] {10.1088/0004-637X/807/2/169},
  \href {https://ui.adsabs.harvard.edu/abs/2015ApJ...807..169A} {807, 169}

\bibitem[\protect\citeauthoryear{{Aharonian}, {Khangulyan}  \&
  {Costamante}}{{Aharonian} et~al.}{2008}]{aharonian2008}
{Aharonian} F.~A.,  {Khangulyan} D.,   {Costamante} L.,  2008, \mn@doi [\mnras]
  {10.1111/j.1365-2966.2008.13315.x}, \href
  {https://ui.adsabs.harvard.edu/abs/2008MNRAS.387.1206A} {387, 1206}

\bibitem[\protect\citeauthoryear{{Albert} et~al.,}{{Albert}
  et~al.}{2007}]{albert2007}
{Albert} J.,  et~al., 2007, \mn@doi [\aap] {10.1051/0004-6361:20078168}, \href
  {https://ui.adsabs.harvard.edu/abs/2007A&A...474..937A} {474, 937}

\bibitem[\protect\citeauthoryear{{Amato}}{{Amato}}{2014}]{amato2014}
{Amato} E.,  2014, \mn@doi [International Journal of Modern Physics D]
  {10.1142/S0218271814300134}, \href
  {https://ui.adsabs.harvard.edu/abs/2014IJMPD..2330013A} {23, 1430013}

\bibitem[\protect\citeauthoryear{{Axford}, {Leer}  \& {Skadron}}{{Axford}
  et~al.}{1977}]{axford1977}
{Axford} W.~I.,  {Leer} E.,   {Skadron} G.,  1977, in International Cosmic Ray
  Conference. p.~132

\bibitem[\protect\citeauthoryear{{Bai} et~al.,}{{Bai} et~al.}{2019}]{LHAASO}
{Bai} X.,  et~al., 2019, arXiv e-prints, \href
  {https://ui.adsabs.harvard.edu/abs/2019arXiv190502773B} {p. arXiv:1905.02773}

\bibitem[\protect\citeauthoryear{{Bell}}{{Bell}}{1978}]{bell1978}
{Bell} A.~R.,  1978, \mn@doi [\mnras] {10.1093/mnras/182.2.147}, \href
  {https://ui.adsabs.harvard.edu/abs/1978MNRAS.182..147B} {182, 147}

\bibitem[\protect\citeauthoryear{{Bell}, {Schure}, {Reville}  \&
  {Giacinti}}{{Bell} et~al.}{2013}]{bell2013}
{Bell} A.~R.,  {Schure} K.~M.,  {Reville} B.,   {Giacinti} G.,  2013, \mn@doi
  [\mnras] {10.1093/mnras/stt179}, \href
  {https://ui.adsabs.harvard.edu/abs/2013MNRAS.431..415B} {431, 415}

\bibitem[\protect\citeauthoryear{{Bell}, {Matthews}  \& {Blundell}}{{Bell}
  et~al.}{2019}]{bell2019}
{Bell} A.~R.,  {Matthews} J.~H.,   {Blundell} K.~M.,  2019, \mn@doi [\mnras]
  {10.1093/mnras/stz1805}, \href
  {https://ui.adsabs.harvard.edu/abs/2019MNRAS.488.2466B} {488, 2466}

\bibitem[\protect\citeauthoryear{{Blandford} \& {Ostriker}}{{Blandford} \&
  {Ostriker}}{1978}]{blandford1978}
{Blandford} R.~D.,  {Ostriker} J.~P.,  1978, \mn@doi [\apjl] {10.1086/182658},
  \href {https://ui.adsabs.harvard.edu/abs/1978ApJ...221L..29B} {221, L29}

\bibitem[\protect\citeauthoryear{{Blasi}}{{Blasi}}{2013}]{blasi2013}
{Blasi} P.,  2013, \mn@doi [\aapr] {10.1007/s00159-013-0070-7}, \href
  {https://ui.adsabs.harvard.edu/abs/2013A&ARv..21...70B} {21, 70}

\bibitem[\protect\citeauthoryear{{Blumenthal}}{{Blumenthal}}{1970}]{Blumenthal70}
{Blumenthal} G.~R.,  1970, \mn@doi [\prd] {10.1103/PhysRevD.1.1596}, \href
  {https://ui.adsabs.harvard.edu/abs/1970PhRvD...1.1596B} {1, 1596}

\bibitem[\protect\citeauthoryear{{Bykov}, {Ellison}, {Gladilin}  \&
  {Osipov}}{{Bykov} et~al.}{2018a}]{bykov2918}
{Bykov} A.~M.,  {Ellison} D.~C.,  {Gladilin} P.~E.,   {Osipov} S.~M.,  2018a,
  \mn@doi [Advances in Space Research] {10.1016/j.asr.2017.05.043}, \href
  {https://ui.adsabs.harvard.edu/abs/2018AdSpR..62.2764B} {62, 2764}

\bibitem[\protect\citeauthoryear{{Bykov}, {Ellison}, {Marcowith}  \&
  {Osipov}}{{Bykov} et~al.}{2018b}]{bykov18}
{Bykov} A.~M.,  {Ellison} D.~C.,  {Marcowith} A.,   {Osipov} S.~M.,  2018b,
  \mn@doi [\ssr] {10.1007/s11214-018-0479-4}, \href
  {https://ui.adsabs.harvard.edu/abs/2018SSRv..214...41B} {214, 41}

\bibitem[\protect\citeauthoryear{{Cardillo}, {Amato}  \& {Blasi}}{{Cardillo}
  et~al.}{2015}]{cardillo2015}
{Cardillo} M.,  {Amato} E.,   {Blasi} P.,  2015, arXiv e-prints, \href
  {https://ui.adsabs.harvard.edu/abs/2015arXiv150706086C} {p. arXiv:1507.06086}

\bibitem[\protect\citeauthoryear{{Celli}, {Morlino}, {Gabici}  \&
  {Aharonian}}{{Celli} et~al.}{2019}]{celli2019}
{Celli} S.,  {Morlino} G.,  {Gabici} S.,   {Aharonian} F.~A.,  2019, \mn@doi
  [\mnras] {10.1093/mnras/stz2897}, \href
  {https://ui.adsabs.harvard.edu/abs/2019MNRAS.490.4317C} {490, 4317}

\bibitem[\protect\citeauthoryear{{Cherenkov Telescope Array Consortium}
  et~al.,}{{Cherenkov Telescope Array Consortium} et~al.}{2019}]{CTAscience}
{Cherenkov Telescope Array Consortium} et~al., 2019, {Science with the
  Cherenkov Telescope Array}, \mn@doi{10.1142/10986.
}

\bibitem[\protect\citeauthoryear{{Chevalier}}{{Chevalier}}{1982}]{chevalier1982}
{Chevalier} R.~A.,  1982, \mn@doi [\apj] {10.1086/160126}, \href
  {https://ui.adsabs.harvard.edu/abs/1982ApJ...258..790C} {258, 790}

\bibitem[\protect\citeauthoryear{{Chevalier} \& {Fransson}}{{Chevalier} \&
  {Fransson}}{1994}]{chevalier1994}
{Chevalier} R.~A.,  {Fransson} C.,  1994, \mn@doi [\apj] {10.1086/173557},
  \href {https://ui.adsabs.harvard.edu/abs/1994ApJ...420..268C} {420, 268}

\bibitem[\protect\citeauthoryear{{Drury}}{{Drury}}{2012}]{drury2012}
{Drury} L. O.~C.,  2012, \mn@doi [Astroparticle Physics]
  {10.1016/j.astropartphys.2012.02.006}, \href
  {https://ui.adsabs.harvard.edu/abs/2012APh....39...52D} {39, 52}

\bibitem[\protect\citeauthoryear{{Drury}, {Aharonian}  \& {Voelk}}{{Drury}
  et~al.}{1994}]{drury1994}
{Drury} L.~O.,  {Aharonian} F.~A.,   {Voelk} H.~J.,  1994, \aap, \href
  {https://ui.adsabs.harvard.edu/abs/1994A&A...287..959D} {287, 959}

\bibitem[\protect\citeauthoryear{{Dubus}}{{Dubus}}{2006}]{dubus2006}
{Dubus} G.,  2006, \mn@doi [\aap] {10.1051/0004-6361:20054233}, \href
  {https://ui.adsabs.harvard.edu/abs/2006A&A...451....9D} {451, 9}

\bibitem[\protect\citeauthoryear{{Fang}, {Metzger}, {Murase}, {Bartos}  \&
  {Kotera}}{{Fang} et~al.}{2019}]{fang2019}
{Fang} K.,  {Metzger} B.~D.,  {Murase} K.,  {Bartos} I.,   {Kotera} K.,  2019,
  \mn@doi [\apj] {10.3847/1538-4357/ab1b72}, \href
  {https://ui.adsabs.harvard.edu/abs/2019ApJ...878...34F} {878, 34}

\bibitem[\protect\citeauthoryear{{Fioretti}, {Bulgarelli}  \&
  {Sch{\"u}ssler}}{{Fioretti} et~al.}{2016}]{fioretti2016}
{Fioretti} V.,  {Bulgarelli} A.,   {Sch{\"u}ssler} F.,  2016, {The Cherenkov
  Telescope array on-site integral sensitivity: observing the Crab}.
p. 99063O, \mn@doi{10.1117/12.2231398}

\bibitem[\protect\citeauthoryear{{Gabici}, {Evoli}, {Gaggero}, {Lipari},
  {Mertsch}, {Orlando}, {Strong}  \& {Vittino}}{{Gabici}
  et~al.}{2019}]{gabici2019}
{Gabici} S.,  {Evoli} C.,  {Gaggero} D.,  {Lipari} P.,  {Mertsch} P.,
  {Orlando} E.,  {Strong} A.,   {Vittino} A.,  2019, arXiv e-prints, \href
  {https://ui.adsabs.harvard.edu/abs/2019arXiv190311584G} {p. arXiv:1903.11584}

\bibitem[\protect\citeauthoryear{{Giacinti} \& {Bell}}{{Giacinti} \&
  {Bell}}{2015}]{Giacinti15}
{Giacinti} G.,  {Bell} A.~R.,  2015, \mn@doi [\mnras] {10.1093/mnras/stv561},
  \href {https://ui.adsabs.harvard.edu/abs/2015MNRAS.449.3693G} {449, 3693}

\bibitem[\protect\citeauthoryear{{Gould} \& {Schr{\'e}der}}{{Gould} \&
  {Schr{\'e}der}}{1966}]{gould1966}
{Gould} R.~J.,  {Schr{\'e}der} G.,  1966, \mn@doi [\prl]
  {10.1103/PhysRevLett.16.252}, \href
  {https://ui.adsabs.harvard.edu/abs/1966PhRvL..16..252G} {16, 252}

\bibitem[\protect\citeauthoryear{{Gould} \& {Schr{\'e}der}}{{Gould} \&
  {Schr{\'e}der}}{1967}]{gould1967}
{Gould} R.~J.,  {Schr{\'e}der} G.~P.,  1967, \mn@doi [Physical Review]
  {10.1103/PhysRev.155.1408}, \href
  {https://ui.adsabs.harvard.edu/abs/1967PhRv..155.1408G} {155, 1408}

\bibitem[\protect\citeauthoryear{{H.~E.~S.~S. Collaboration}
  et~al.,}{{H.~E.~S.~S. Collaboration} et~al.}{2018}]{HESSSNR}
{H.~E.~S.~S. Collaboration} et~al., 2018, \mn@doi [\aap]
  {10.1051/0004-6361/201732125}, \href
  {https://ui.adsabs.harvard.edu/abs/2018A&A...612A...3H} {612, A3}

\bibitem[\protect\citeauthoryear{{H.~E.~S.~S. Collaboration}
  et~al.,}{{H.~E.~S.~S. Collaboration} et~al.}{2019}]{Hess19}
{H.~E.~S.~S. Collaboration} et~al., 2019, \mn@doi [\aap]
  {10.1051/0004-6361/201935242}, \href
  {https://ui.adsabs.harvard.edu/abs/2019A&A...626A..57H} {626, A57}

\bibitem[\protect\citeauthoryear{{Kirk}, {Duffy}  \& {Ball}}{{Kirk}
  et~al.}{1995}]{kirk1995}
{Kirk} J.~G.,  {Duffy} P.,   {Ball} L.,  1995, \aap, \href
  {https://ui.adsabs.harvard.edu/abs/1995A&A...293L..37K} {293, L37}

\bibitem[\protect\citeauthoryear{{Krymskii}}{{Krymskii}}{1977}]{krymskii1977}
{Krymskii} G.~F.,  1977, Akademiia Nauk SSSR Doklady, \href
  {https://ui.adsabs.harvard.edu/abs/1977DoSSR.234.1306K} {234, 1306}

\bibitem[\protect\citeauthoryear{{Lewis} et~al.,}{{Lewis}
  et~al.}{1994}]{lewis1994}
{Lewis} J.~R.,  et~al., 1994, \mn@doi [\mnras] {10.1093/mnras/266.1.L27}, \href
  {https://ui.adsabs.harvard.edu/abs/1994MNRAS.266L..27L} {266, L27}

\bibitem[\protect\citeauthoryear{{Malkov} \& {Aharonian}}{{Malkov} \&
  {Aharonian}}{2019}]{malkov2019}
{Malkov} M.~A.,  {Aharonian} F.~A.,  2019, \mn@doi [\apj]
  {10.3847/1538-4357/ab2c01}, \href
  {https://ui.adsabs.harvard.edu/abs/2019ApJ...881....2M} {881, 2}

\bibitem[\protect\citeauthoryear{{Marcowith}, {Renaud}, {Dwarkadas}  \&
  {Tatischeff}}{{Marcowith} et~al.}{2014}]{marcowith2014}
{Marcowith} A.,  {Renaud} M.,  {Dwarkadas} V.,   {Tatischeff} V.,  2014,
  \mn@doi [Nuclear Physics B Proceedings Supplements]
  {10.1016/j.nuclphysbps.2014.10.011}, \href
  {https://ui.adsabs.harvard.edu/abs/2014NuPhS.256...94M} {256, 94}

\bibitem[\protect\citeauthoryear{{Marcowith}, {Dwarkadas}, {Renaud},
  {Tatischeff}  \& {Giacinti}}{{Marcowith} et~al.}{2018}]{marcowith2018}
{Marcowith} A.,  {Dwarkadas} V.~V.,  {Renaud} M.,  {Tatischeff} V.,
  {Giacinti} G.,  2018, \mn@doi [\mnras] {10.1093/mnras/sty1743}, \href
  {https://ui.adsabs.harvard.edu/abs/2018MNRAS.479.4470M} {479, 4470}

\bibitem[\protect\citeauthoryear{{Maund}, {Smartt}, {Kudritzki},
  {Podsiadlowski}  \& {Gilmore}}{{Maund} et~al.}{2004}]{maund2004}
{Maund} J.~R.,  {Smartt} S.~J.,  {Kudritzki} R.~P.,  {Podsiadlowski} P.,
  {Gilmore} G.~F.,  2004, \mn@doi [\nat] {10.1038/nature02161}, \href
  {https://ui.adsabs.harvard.edu/abs/2004Natur.427..129M} {427, 129}

\bibitem[\protect\citeauthoryear{{Moriya}, {Maeda}, {Taddia}, {Sollerman},
  {Blinnikov}  \& {Sorokina}}{{Moriya} et~al.}{2014}]{moriya2014}
{Moriya} T.~J.,  {Maeda} K.,  {Taddia} F.,  {Sollerman} J.,  {Blinnikov} S.~I.,
    {Sorokina} E.~I.,  2014, \mn@doi [\mnras] {10.1093/mnras/stu163}, \href
  {https://ui.adsabs.harvard.edu/abs/2014MNRAS.439.2917M} {439, 2917}

\bibitem[\protect\citeauthoryear{{Murase}, {Thompson}  \& {Ofek}}{{Murase}
  et~al.}{2014}]{murase2014}
{Murase} K.,  {Thompson} T.~A.,   {Ofek} E.~O.,  2014, \mn@doi [\mnras]
  {10.1093/mnras/stu384}, \href
  {https://ui.adsabs.harvard.edu/abs/2014MNRAS.440.2528M} {440, 2528}

\bibitem[\protect\citeauthoryear{{Murase}, {Franckowiak}, {Maeda}, {Margutti}
  \& {Beacom}}{{Murase} et~al.}{2019}]{murase2019}
{Murase} K.,  {Franckowiak} A.,  {Maeda} K.,  {Margutti} R.,   {Beacom} J.~F.,
  2019, \mn@doi [\apj] {10.3847/1538-4357/ab0422}, \href
  {https://ui.adsabs.harvard.edu/abs/2019ApJ...874...80M} {874, 80}

\bibitem[\protect\citeauthoryear{{Petropoulou}, {Coenders}, {Vasilopoulos},
  {Kamble}  \& {Sironi}}{{Petropoulou} et~al.}{2017}]{petropoulou17}
{Petropoulou} M.,  {Coenders} S.,  {Vasilopoulos} G.,  {Kamble} A.,   {Sironi}
  L.,  2017, \mn@doi [\mnras] {10.1093/mnras/stx1251}, \href
  {https://ui.adsabs.harvard.edu/abs/2017MNRAS.470.1881P} {470, 1881}

\bibitem[\protect\citeauthoryear{{Ripero} et~al.,}{{Ripero}
  et~al.}{1993}]{ripero1993}
{Ripero} J.,  et~al., 1993, \iaucirc, \href
  {https://ui.adsabs.harvard.edu/abs/1993IAUC.5731....1R} {5731, 1}

\bibitem[\protect\citeauthoryear{{Schure} \& {Bell}}{{Schure} \&
  {Bell}}{2014}]{schure2014}
{Schure} K.~M.,  {Bell} A.~R.,  2014, \mn@doi [\mnras] {10.1093/mnras/stt2089},
  \href {https://ui.adsabs.harvard.edu/abs/2014MNRAS.437.2802S} {437, 2802}

\bibitem[\protect\citeauthoryear{{Smartt}}{{Smartt}}{2009}]{smartt2009}
{Smartt} S.~J.,  2009, \mn@doi [\araa] {10.1146/annurev-astro-082708-101737},
  \href {https://ui.adsabs.harvard.edu/abs/2009ARA&A..47...63S} {47, 63}

\bibitem[\protect\citeauthoryear{{Tamborra} \& {Murase}}{{Tamborra} \&
  {Murase}}{2018}]{tamborra18}
{Tamborra} I.,  {Murase} K.,  2018, \mn@doi [\ssr] {10.1007/s11214-018-0468-7},
  \href {https://ui.adsabs.harvard.edu/abs/2018SSRv..214...31T} {214, 31}

\bibitem[\protect\citeauthoryear{{Tatischeff}}{{Tatischeff}}{2009}]{tatischeff2009}
{Tatischeff} V.,  2009, \mn@doi [\aap] {10.1051/0004-6361/200811511}, \href
  {https://ui.adsabs.harvard.edu/abs/2009A&A...499..191T} {499, 191}

\bibitem[\protect\citeauthoryear{{Wang}, {Huang}  \& {Li}}{{Wang}
  et~al.}{2019}]{wang2019}
{Wang} K.,  {Huang} T.-Q.,   {Li} Z.,  2019, \mn@doi [\apj]
  {10.3847/1538-4357/aaffd9}, \href
  {https://ui.adsabs.harvard.edu/abs/2019ApJ...872..157W} {872, 157}

\bibitem[\protect\citeauthoryear{{Zirakashvili} \& {Ptuskin}}{{Zirakashvili} \&
  {Ptuskin}}{2016}]{zirakashvili2016}
{Zirakashvili} V.~N.,  {Ptuskin} V.~S.,  2016, \mn@doi [Astroparticle Physics]
  {10.1016/j.astropartphys.2016.02.004}, \href
  {https://ui.adsabs.harvard.edu/abs/2016APh....78...28Z} {78, 28}

\makeatother
\end{thebibliography}




\appendix

\section{Derivation of $\psi_{0,\rm min}$}
\label{app:A}
$\psi_{0,\rm min}$ corresponds to the limit below which gamma--ray photons are absorbed because they have to cross the photosphere. This condition can be written: 
\begin{equation}
    R_{\rm sh}(t) \sin (\psi_{0,\rm min}(t)) = R_{\rm ph}(t+t_+)
\end{equation}
with $t_+>0$ the unknown to be found. Geometrically, we additionally get: 
\begin{equation}
    t_+= \frac{R_{\rm sh}(t)}{c \cos(\psi_{0,\rm min}(t))}
\end{equation}
and can calculate at each time step $t$ the value of $\psi_{0,\rm min}$ between 0 and $\pi/2$. Numerically, it is convenient to introduce $\mu = \cos(\psi_{0,\rm min})$, taking values between 0 and 1, and search for the greatest value of $\mu$ satisfying: 
\begin{equation}
    R^2_{\rm sh}(t) (1-\mu^2)= R^2_{\rm ph}(t+ R_{\rm sh}(t)/(c\mu))~.
\end{equation}

\bsp	
\label{lastpage}
\end{document}